\documentclass[sigconf,screen]{acmart}
  \AtBeginDocument{%
    }

  \usepackage{algorithm}
  \usepackage{algpseudocodex}
  \usepackage{graphicx}
  \usepackage{textcomp}
  \usepackage{xcolor}
  \usepackage{float}
  \usepackage{subcaption}
  \usepackage{booktabs}
  \usepackage{amsmath}
  \usepackage{multirow}
  \usepackage{tikz}
  \usepackage{booktabs}
  \usepackage{enumitem}
  \usepackage{hyperref}
  \usepackage[para]{footmisc}
  \usepackage{pifont}
  \algnewcommand{\IfReturn}[2]{%
    \If{#1} \textbf{return} #2; \EndIf%
  }
  \setcopyright{acmlicensed}
  \copyrightyear{2025}
  \acmYear{2025}
  \acmDOI{10.1145/3770743.3803891}
  \acmConference[DAC '26]{The Chips to Systems Conference}{July 26--29,
    2026}{Long Beach, CA}
  \acmISBN{XXX-X-XXXX-XXXX-X/2026/07}

  \acmSubmissionID{4}

\begin{document}

  \title[Warp-STAR]{Warp-STAR: High‑performance, Differentiable GPU-Accelerated \textbf{\underline{S}}tatic \textbf{\underline{T}}iming \textbf{\underline{A}}nalysis through Wa\textbf{\underline{r}}p-oriented Parallel Orchestration}


  \author{En-Ming Huang}
  \email{r13922078@csie.ntu.edu.tw}
  \affiliation{%
    \institution{National Taiwan University}
    \city{Taipei}
    \country{Taiwan}
  }
  \author{Shih-Hao Hung}
  \email{hungsh@csie.ntu.edu.tw}
  \affiliation{%
    \institution{National Taiwan University}
    \city{Taipei}
    \country{Taiwan}
  }
  



  \begin{abstract}

Static timing analysis (STA) is crucial for Electronic Design Automation (EDA) flows but remains a computational bottleneck. While existing GPU-based STA engines are faster than CPU, they suffer from inefficiencies, particularly intra-warp load imbalance caused by irregular circuit graphs. This paper introduces Warp-STAR, a novel GPU-accelerated STA engine that eliminates this imbalance by orchestrating parallel computations at the warp level. This approach achieves a 2.4X speedup over previous state-of-the-art (SoTA) GPU-based STA. When integrated into a timing-driven global placement framework, Warp-STAR delivers a 1.7X speedup over SoTA frameworks. The method also proves effective for differentiable gradient analysis with minimal overhead.

\end{abstract}

  \keywords{Static Timing Analysis, GPU Acceleration, Irregular Graphs, Intra-Warp Load Imbalance}

  \maketitle

  \section{Introduction}
\label{sec:intro}

Static timing analysis (STA) is widely adopted across various stages of the EDA design flow, providing critical insights into timing optimization and correction~\cite{2019openroad,OpenSTA,2015TAU,2015ICCAD,2023DP4}. 
For instance, timing-driven global placement frameworks~\cite{2023DP4,2025EfficientTDP,guo2022difftdp,2025Xplace3,lu2025insta} leverage STA engines to optimize cell delays alongside standard metrics like half-perimeter wire length (HPWL).
Nonetheless, these frameworks require hundreds of iterations of STA, resulting in significant runtime overhead. DreamPlace 4.0~\cite{2023DP4}, a timing-driven global placement (GP) framework, reports that the CPU-based STA engine~\cite{Huang2021OpenTimerV2} accounts for 51.3\% of the total runtime during the GP phase in a design with 2.5 million pins, limiting invocations to once per 15 training iterations~\cite{2023DP4}, making STA efficiency a pressing challenge.

STA engines are accelerated through either multithreading or GPU-based parallelism. OpenTimer (OT)~\cite{Huang2021OpenTimerV2} integrates Taskflow~\cite{huang2022taskflow} to enhance performance on multi-core CPUs. GPU-based engines~\cite{Wang2014CASTA,Guo2020GPUSTA, Guo2023CPUGPUSTA,lu2025insta} achieve higher performance, contributing 3.7X speedup compared to OT running on a 40-core CPU~\cite{Guo2020GPUSTA}. However, these works omit their low execution efficiency on the GPU because circuits are irregular graphs which are inefficient for GPU processing.

We found that GPU-based STA engines~\cite{Wang2014CASTA,Guo2020GPUSTA, Guo2023CPUGPUSTA} map each net to individual GPU threads (CUDA threads), leading to the intra-warp load imbalance issue. Each thread performs computation on its own net, and since each net has a different number of pins, the number of iterations differs across threads. Due to the design of the GPU, threads in the same warp are executed in lockstep, leading to underutilized resources.
Beyond intra-warp load imbalance, differentiable STA~\cite{guo2022difftdp,lu2025insta} introduces a further overhead: existing frameworks execute gradient backward pass \emph{sequentially after} the full STA pipeline, leaving the GPU idle between synchronizations, reducing GPU utilization and adding substantial per-iteration cost.

In this work, we present Warp-STAR, an efficient, differentiable GPU-accelerated \textbf{\underline{STA}} through wa\textbf{\underline{r}}p-oriented parallel orchestration to eliminate intra-warp load imbalances. Unlike previous works that map nets to individual threads, workloads (i.e., the pins within a net) are processed by the unit of warp. As a result, we enhance the utilization of CUDA warps, achieving an average of 2.4X speedup in comparison with existing GPU-based STA~\cite{Wang2014CASTA,Guo2020GPUSTA,Guo2023CPUGPUSTA}, and 162X speedup against CPU-based STA, OT~\cite{Huang2021OpenTimerV2} on the 2015 ICCAD contest dataset~\cite{2015ICCAD}. Additionally, we incorporate our STA engine into a timing-driven GP framework based on Xplace 3.0~\cite{2025Xplace3}, delivering state-of-the-art execution time and high quality of timing slack compared with recent SoTA works~\cite{2023DP4, 2025EfficientTDP, 2025Xplace3,lu2025insta}. We further reduce the differentiable backward-pass overhead through \textit{operation fusion}, by allowing concurrent STA computations and gradient computations via CUDA streams, achieving superior runtime over INSTA~\cite{lu2025insta}.
Our main contributions are threefold:
\begin{enumerate}[noitemsep,topsep=0pt,leftmargin=*]
    \item \textbf{Tackling Intra-Warp Load Imbalance}: We identify and resolve the intra-warp load imbalance problem in GPU-based STA engines through our proposed Warp-STAR.

    \item \textbf{Operation Fusion for Differentiable STA}: We reduce the sequential backward-pass overhead of differentiable STA by overlapping gradient computations with STA's slack propagation, enhancing GPU utilization in gradient analysis.

    \item \textbf{Versatile Performance with Warp-STAR}: Our optimized STA engine provides improved performance across applications, including existing GP frameworks and gradient analysis.
\end{enumerate}
  \section{Preliminary}
\label{sec:background}

In this section, we introduce the background knowledge of STA and GPU architecture along with the problem of intra-warp load imbalance in graph processing.

\subsection{Static Timing Analysis}
\begin{figure}[t]
    \centering
    \begin{subfigure}[t]{0.23\textwidth}
        \centering
        \raisebox{0pt}[\height][\depth]{\includegraphics[width=\linewidth,page=1]{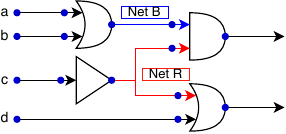}}
        \caption{An illustration of a circuit design with its components: pins, cells, and nets.}
        \label{fig:circuits}
    \end{subfigure}
    \hfill
    \begin{subfigure}[t]{0.20\textwidth}
        \centering
        \raisebox{0pt}[\height][\depth]{\includegraphics[width=\linewidth,page=2]{figures/img/2.circuits.pdf}}
        \caption{Visualization of the \textit{levelization} process, where nets are grouped into levels based on their dependencies.}
        \label{fig:levelization}
    \end{subfigure}
    \vspace{-1.2em}
    \caption{Circuit components and \textit{levelization} in STA. (a) The components of a circuit design. (b) The \textit{levelization} process.}
    \label{fig:circuits-levelization}
    \vspace{-1.5em}
\end{figure}

In STA, the problem is formulated as a Directed Acyclic Graph (DAG) and comprises three main components: pins, cells, and nets. Figure~\ref{fig:circuits} illustrates these components. Each gate (e.g., AND, OR, BUFFER, etc.) is represented as a cell, while the blue dots associated with the cells are the pins, representing their input or output ports. Wire connections are represented by arrows. Wires of the same color, such as the red wires, are grouped as a net. A net has one input (driver) pin and multiple output pins to represent a fanout. The signal arrival time (AT) of each cell can be computed through a forward Breadth-First Search (BFS) propagation by taking the maximum AT of its preceding cells and adding the delay of wires and itself. Once all ATs are computed, slacks can be derived from a backward propagation using another BFS operation by subtracting the calculated AT from the required arrival time (RAT).

The method of calculating STA in parallel is commonly achieved through the following four stages:
\textbf{(1) RC net delay computation}: Computes the delay of wires in each net and can be done in parallel. \textbf{(2) Data dependency identification (\textit{levelization})}: Groups independent nets together, which is crucial for calculating cell AT propagation since data dependency relationships exist between cells in the DAG. As Figure~\ref{fig:levelization} shows, the circuit in Figure~\ref{fig:circuits} can be formulated into three levels of nets, where the pins belonging to level $i$ require AT computed in level $i-1$. \textbf{(3) Forward propagation for cell AT}: This stage computes the cell delay along with its AT level by level to ensure data correctness. However, computations within each level can be executed in parallel since no data dependency exists. \textbf{(4) Backward propagation for cell slacks}: Computes cell slacks from the opposite direction, starting from the output pins of the last level.
It can be noted that for flows requiring multiple STA executions (e.g., timing-driven GP), stage (2) needs to be computed only once because the design's netlist is not modified. Therefore, the primary bottlenecks are the RC net delay computation, as well as the forward and backward propagation stages.

\subsection{GPU Architecture and Irregular Graph Processing}
Efficient execution of graph algorithms, such as BFS, on GPUs remains challenging due to irregular graph structures and the GPU's Single-Instruction Multiple-Threads (SIMT) execution model~\cite{2024WER}.
In the SIMT execution model, threads within a warp operate in lockstep. A group of threads are ultimately mapped to a Single-Instruction Multiple-Data (SIMD) execution unit that fetches and attempts to execute the same instruction simultaneously. Nonetheless, graph algorithms often exhibit control-flow divergence due to conditional branches or workload imbalance within a warp (e.g., threads iterating over neighbors with vastly different degrees). To maintain correctness during such divergence, SIMT masks are employed and maintained by the hardware. These masks disable SIMD lanes corresponding to threads that do not follow the currently active execution path. While this ensures correctness, it critically results in idled SIMD lanes and unnecessarily extends the warp's overall completion time. This problem is intensified by typical warp sizes: NVIDIA's GPUs group threads into warps of 32, while AMD's CDNA GPUs use warps comprising 64 threads.
\begin{figure}[t]
    \centering
    \includegraphics[width=0.92\linewidth]{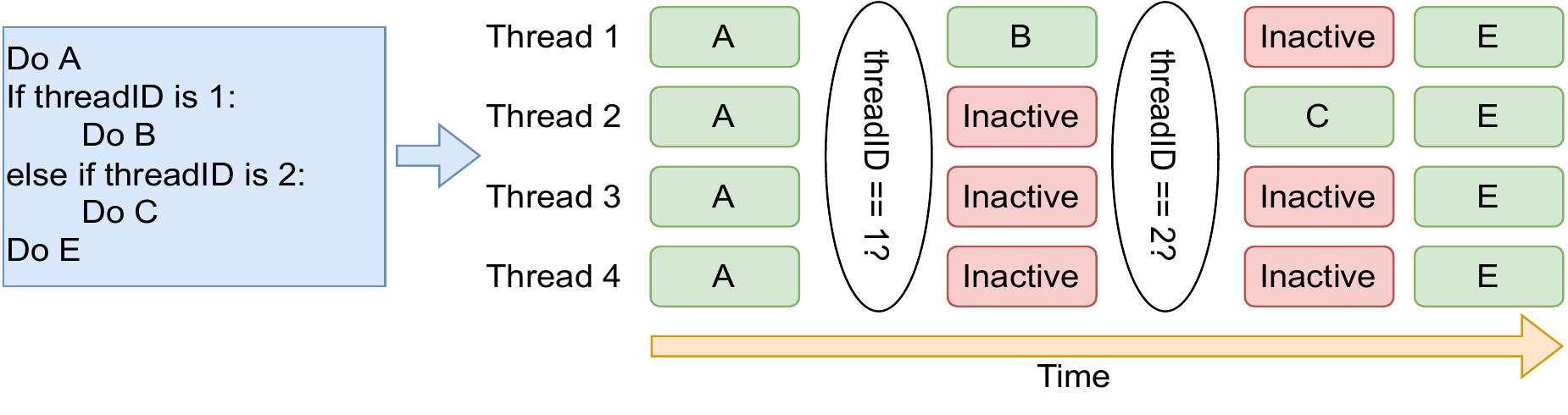}
    \vspace{-1.3em}
    \caption{\small Illustration of thread divergence in a 4-threaded warp.}
    \label{fig:thread_div}
    \vspace{-1.5em}
\end{figure}

Figure~\ref{fig:thread_div} illustrates a common challenge: branch divergence within a 4-threaded warp. This phenomenon leads to underutilized SIMD units when threads within the same warp take different execution paths. A similar inefficiency occurs when threads have varying numbers of iterations within loops.

To maximize computational throughput and hide latency (e.g., memory requests), GPUs dynamically schedule multiple warps on their cores (i.e., stream-multiprocessors). When a warp's workload finishes or is stalled, the GPU scheduler can switch to other warps, even if some warps are still processing lengthy tasks. Consequently, inter-warp load imbalance is less critical than intra-warp load imbalance, as the hardware scheduler efficiently switches between warps to maintain high utilization.
Recent studies have proposed approaches such as graph transformations~\cite{2018Tigr}, workload rescheduling~\cite{2015simdcte2,2016CTE}, and hardware-assisted techniques~\cite{2024WER,Segura2019SCU} to improve SIMD lane utilization for general graph algorithms. These works all aim to address intra-warp load imbalances.

  \section{Methodology}
\label{sec:methodology}

\begin{figure}[t]
    \centering
    \includegraphics[width=0.94\linewidth]{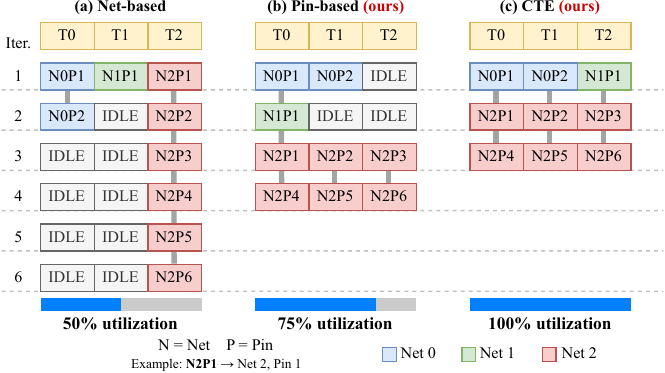}
    \vspace{-1.1em}
    \caption{Comparison of task scheduling strategies for threads T0$\sim$T2. \textbf{(a) Net based scheme}: each thread processes an entire net. \textbf{(b) Pin based scheme}: each thread processes individual pins, allowing for higher parallelism. \textbf{(c) Collaborative Task Engagement (CTE)}, which dynamically reschedules workloads within the thread block to maximize utilization.}
    \label{fig:workloadcmp}
    \vspace{-1.3em}
\end{figure}

Warp-STAR introduces two methods to eliminate intra-warp inefficiencies through warp-centric parallelism: a pin-based scheme and Collaborative Task Engagement (CTE)~\cite{2016CTE,2015simdcte2}. In contrast to prior GPU-based STA engines~\cite{Wang2014CASTA,Guo2020GPUSTA,Guo2023CPUGPUSTA} that employed a net-based scheme, our approach distributes the workloads of each net (i.e., its pins) across threads to enhance SIMD lane utilization.

Figure~\ref{fig:workloadcmp} illustrates the key difference: in a traditional net-based scheme, employed by previous GPU-based STAs, entire nets are mapped to individual threads. This can lead to underutilization of other threads (or SIMD lanes) if a net has a large number of pins (e.g., as shown for thread 2 in Figure~\ref{fig:workloadcmp}(a)). Conversely, our pin-based scheme enables threads to process workloads at the granularity of individual pins, allowing large nets to be processed in parallel for reduced execution time. Furthermore, we introduce CTE, which dynamically reschedules workloads within the thread block before computation to achieve maximum utilization. Although CTE enables the best SIMD lane utilization, it incurs overhead for workload rescheduling and does not yield optimal performance compared to the pin-based scheme. Nevertheless, both methods provide performance acceleration compared to the baseline GPU-based STA, as detailed comparisons examined in Section~\ref{sec:experiments}.

In this section, we first introduce the parallel algorithm of our work and the procedures to deal with race conditions across threads. Next, we showcase the \textit{operation fusion} for efficient gradient computation. Finally, we incorporate Warp-STAR into a global placement framework for efficient timing-driven GP optimization.

\subsection{Algorithm of Warp-Centric STA Computation}
\begin{figure}[t]
\vspace{-1.5em}
\begin{minipage}{\columnwidth}
    \begin{algorithm}[H]
    \caption{RC Net Delay in the Pin-based Scheme}
    \label{alg:rcnetpin}
    \begin{algorithmic}[1] 
    \State \textbf{CPU:}
    \State N\_blocks $\gets \lceil$ num\_nets / NETS\_PER\_BLOCK $\rceil$
    \State \Call{RCNet}{$<$dim3(4,8,NETS\_PER\_BLOCK), N\_blocks$>$}

    \State \textbf{GPU:}
    \Function{RCNet}{}
    \State \textbf{Input:} N as \# of nets, P as \# of pins
    \State \textbf{Input:} netlist[0 \dots P], the netlist in Compressed Sparse Row (CSR) format
    \State \textbf{Input:} netlist\_ind[0 \dots N], the indexes of each net
    \State \textbf{Output:} load[0 \dots P], delay[0 \dots P], Impulse[0 \dots P]

    \State netID $\gets$ \text{blockIdx.x} $\cdot$ \text{blockDim.z} + \text{threadIdx.z};
    \State localTID $\gets$ \text{threadIdx.y};
    \State condID $\gets$ \text{threadIdx.x}; \Comment{Shared memory}
    \State \textbf{shared} float shmem[4][8][NETS\_PER\_BLOCK]; 
    \IfReturn{netID $\ge$ N}{}
    \State netRoot $\gets$ netlist\_ind[netID];
    \State netStart $\gets$ netlist\_ind[netID] + localTID;
    \State netEnd $\gets$ netlist\_ind[netID+1];

    \LComment{Parallel calculation for pin loads}
    \State shmem[condID][localTID][threadIdx.z] $\gets$ 0;
    \For{i = netStart; i $<$ netEnd; i += blockDim.y}
        \State load[i $\cdot$ 4 + condID] $\gets$ \text{calc\_load()};
        \State shmem[condID][localTID][threadIdx.z] += load[i $\cdot$ 4 + condID];
    \EndFor
    \State \_\_syncwarp()

    \LComment{Parallel reduction for root load}
    \For{i = 1; i $<$ blockDim.y; i *= 2}
        \If{localTID $\bmod$ (i $\cdot$ 2) == 0}
            \State shmem[condID][localTID][threadIdx.z] += shmem[condID][localTID + i][threadIdx.z];
        \EndIf
        \State \_\_syncwarp()
    \EndFor
    \If{localTID == 0}
        \State load[netRoot $\cdot$ 4 + condID] $\gets$ shmem[condID][0][threadIdx.z];
    \EndIf

    \LComment{Parallel calculation for pin delays}
    \For{i = netStart; i $<$ netEnd; i += blockDim.y}
        \State delay[i $\cdot$ 4 + condID] $\gets$ res[i+condID] $\cdot$ load[i+condID];
    \EndFor

    \LComment{Parallel calculation for pin impulses}
    \For{i = netStart; i $<$ netEnd; i += blockDim.y}
        \State impulse[i $\cdot$ 4 + condID] $\gets \sqrt{\dots}$;
    \EndFor
    \EndFunction
    \end{algorithmic}
    \end{algorithm}
\end{minipage}
\vspace{-1.8em}
\end{figure}

\begin{figure}[t]
\vspace{-1.3em}
\begin{minipage}{\columnwidth}
    \begin{algorithm}[H]
    \caption{RC Net Delay in the CTE~\cite{2016CTE,2015simdcte2} Scheme}
    \label{alg:rcnetcte}
    \begin{algorithmic}[1] 
    \State \textbf{CPU:}
    \State N\_blocks $\gets \lceil$ num\_nets / NETS\_PER\_BLOCK $\rceil$
    \State \Call{RCNet\_CTE}{$<$NETS\_PER\_BLOCK, N\_blocks$>$}

    \State \textbf{GPU:}
    \Function{RCNet\_CTE}{}
    \State netID $\gets$ \text{blockIdx.x} $\cdot$ \text{blockDim.x} + \text{threadIdx.x};
    \State \textbf{shared} int workload\_prefix[NETS\_PER\_BLOCK];

    \State netStart $\gets$ netlist\_ind[netID]+1;
    \State netEnd $\gets$ netlist\_ind[netID+1];
    \State workload\_prefix[threadIdx.x] $\gets$ (netEnd - netStart)*4;
    \LComment{Parallel prefix sum (Work-Efficient Parallel Scan) on workload\_prefix}
    \State $\dots$ 
    \State total\_workloads $\gets$ workload\_prefix[blockDim.x-1];
    \LComment{Calculate}
    \For{taskID = threadIdx.x; taskID $<$ total\_workloads; taskID += blockDim.x}
        \State taskNetID $\gets$ \text{lower\_bound(workload\_prefix, taskID)};
        \State taskNetRoot $\gets$ netlist\_ind[taskNetID];
        \State taskPinID $\gets$ {\small(taskID - workload\_prefix[taskNetID]) / 4;}
        \State taskCondID $\gets$ {\small(taskID - workload\_prefix[taskNetID]) \% 4;}
        \State \textit{Do work}\dots
    \EndFor
    \EndFunction
    \end{algorithmic}
    \end{algorithm}
\end{minipage}
\vspace{-1.5em}
\end{figure}
Our algorithm divides the STA computation into two main phases: (1) net delay calculation and (2) arrival time (AT) propagation. This separation is key for performance, as the net delay calculations are highly parallelizable, while the AT propagation through cells is inherently sequential due to data dependencies that are managed by the process of \textit{levelization}~\cite{Guo2020GPUSTA}. First, we describe the algorithm of \textit{RC net delay calculation} in a warp-centric manner, followed by the \textit{AT propagation and slack computation} process.

\subsubsection{\textbf{RC Net Delay Computation}}

The RC delay of a single net is estimated using the Elmore delay model, described by Equations~\ref{eq:load}$\sim$\ref{eq:impulse}~\cite{2015TAU}:
\begin{equation}
    Load(u) = Cap(u) + \textstyle\sum_{v \in Fanout(u)} Load(v)
    \label{eq:load}
\end{equation}
\begin{equation}
    Delay(u) = Res\big(fa(u) \rightarrow u\big) \cdot Load(u)
    \label{eq:delay}
\end{equation}
\begin{equation}
    Impulse(u) = \sqrt{2 \cdot Res(u) \cdot Cap(u) \cdot Delay(u) - Delay(u)^2}
    \label{eq:impulse}
\end{equation}

Here, $fa(u)$ refers to the parent pin of $u$, specifically the input pin of its associated net.
These equations are applied to each pin ($u$) within a net, inherently revealing the load imbalance problem across nets due to varying numbers of pins. As a result, instead of mapping each net to individual CUDA threads, in Warp-STAR, the workload of a net is processed by a warp. Note that the summation operator in Equation~\ref{eq:load} highlights that the parent pin requires the sum of its output pin loads. In our implementation, we employ \textit{parallel reduction} to avoid race conditions and atomic operations.

Algorithm~\ref{alg:rcnetpin} presents the pseudo code for our pin-based scheme. Threads are grouped into a three-dimensional format: (x=4, y=8, z=NETS\_PER\_BLOCK). The X-dimension (4 threads) is dedicated to processing a single pin across four timing conditions (early/late and fall/rise). The Y-dimension (8 threads) forms a warp of 32 threads. Finally, the Z-dimension is utilized to form a larger thread block, thereby reducing hardware scheduling overhead due to the large amount of thread blocks. We leverage shared memory, a scratchpad memory from the L1 cache, for storing shared data across threads and facilitating parallel reduction computations.

Additionally, we present the CTE algorithm in Algorithm~\ref{alg:rcnetcte}. CTE optimizes workload distribution within a thread block by reassigning tasks dynamically. First, each CUDA thread obtains its assigned net and its workloads (i.e., pins). These workloads are combined to form a unified task pool using a parallel prefix sum operation, which gathers the jobs. The parallel prefix sum operation is based on the work-efficient data-parallel scanning algorithm~\cite{BlellochTR90}, consisting of two stages: reduce and down-sweep. This method is efficient for scenarios where parallel threads, such as those from different warps, are not perfectly synchronized. Therefore, this allows for workload rescheduling across warps, not just within a single warp, thereby reducing inter-warp load imbalance. Once the workloads are organized into the prefix sum array, computations begin. Each thread retrieves a single task (i.e., a pin) through striding. Subsequently, it calculates its belonging net by performing a binary search on the prefix sum array to find the maximum index, whose value is less than or equal to the task index. This operation is also referred to as the \texttt{std::lower\_bound} in the C++ standard library. After the workload is retrieved, the \textit{Do work} continues to the same RC net delay computation as Line $19\sim36$ in Algorithm~\ref{alg:rcnetpin}.

\subsubsection{\textbf{Cell arrival time and Slack propagation}}
The propagation of arrival times (ATs) across cells requires updating the output pin's AT based on the input pins' ATs and the calculated cell delays through LUT interpolations. For each input-to-output path within a cell, we determine the cell arc delay by interpolating values from a look-up table (LUT). The cell's late-mode output AT is then computed by taking the maximum over all input pins of the sum of the input pin's AT and its corresponding cell arc delay. Conversely, the early-mode AT is derived using the minimum.

The calculation of pin and cell slacks is performed by subtracting the AT from the required arrival time (RAT), which is computed via a backward propagation pass, starting from the circuit's primary outputs. Therefore, in Warp-STAR, ATs and slacks are calculated in two distinct stages: a forward pass for ATs and a backward pass for slacks. We apply the same parallelization strategies used in the \textit{RC Net Delay Computation}, the pin-based scheme or the CTE scheme, to accelerate the AT propagation and slack computation.

To ensure the correct execution order of AT propagation for parallel processing, nets are organized into a sequential order, consisting of multiple levels, through the process of \textit{levelization}~\cite{Wang2014CASTA}. Independent nets are grouped into the same level and can be processed in parallel, while dependent nets are placed in different levels to manage their data dependencies, as shown in Figure~\ref{fig:levelization}. Our implementation follows the method presented in \cite{Guo2020GPUSTA}, where each level is described by a CUDA kernel function.

\subsection{\textit{Operation Fusion} for Differentiable STA}
In traditional STA, the arrival time at a cell's output pin is determined using a maximum operation over the ATs from all input paths. However, it is unsuitable for gradient analysis because only the single largest input value contributes to the maximum, preventing the gradient from being effectively shared among other pins. To overcome this limitation, many works have adopted the LogSumExp (LSE) operation as a differentiable replacement~\cite{guo2022difftdp,lu2025insta}, which can be expressed as Equation~\ref{eq:lse}
\begin{equation}
    \label{eq:lse}
    y=c+\gamma\log\textstyle\sum_{i=1}^{n}exp\left(\frac{x_i-c}{\gamma}\right)
\end{equation}
,where $x_i$ represents the values intended for the LSE operation, $c$ is the maximum value among the $x_i$ values, and $\gamma$ is a hyper-parameter that controls the smoothness of the output. The term $c$ is crucial for preventing floating-point overflow when  $e^{x_i/\gamma}$ are large enough to exceed the floating point's range.

\begin{figure}[t]
\centering
\includegraphics[width=0.75\linewidth]{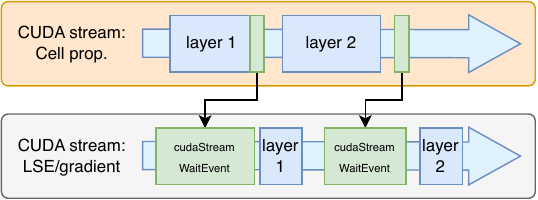}
\vspace{-1em}
\caption{Overlapped execution of AT propagation and gradient computation in Warp-STAR. CUDA events are employed, allowing concurrent execution while ensuring correctness.}
\label{fig:gradient_cudastream}
\vspace{-1.5em}
\end{figure}

To obtain gradients, we calculate pin ATs using both the maximum and LSE operations, the former for traditional timing calculations and the latter for differentiable computations. The LSE operations follow the same data dependency relationships as AT propagation, thus leveraging the same \textit{levelization} results. However, since the LSE operation depends on cell arc delays (derived from LUT interpolations during the AT propagation stage), it must be scheduled after the cell delay for a given level has been computed.

We further propose a tight integration with the original STA through \textit{\textbf{operation fusion}}, which overlaps slack propagation and gradient computations using CUDA streams. Instead of calculating LSE and gradient after the original STA, which consists of the \textit{RC net delay} and the \textit{AT propagation} stages, we demonstrate that it can be performed concurrently with the AT propagation stage. This strategy maximizes GPU's computing throughput and reduces the overhead associated with gradient computations.
Our key observation is that separating gradient calculations into distinct forward and backward passes, a common practice when integrating automatic differentiation frameworks like PyTorch (e.g., INSTA~\cite{lu2025insta} requires a manual \texttt{insta.tns.backward()} invocation after the slack calculation) is unnecessary in our context. This is primarily for two reasons: (1) calculating cell slacks inherently involves a backward propagation step within the timing analysis; and (2) most implementations derive the gradient directly from the loss function of total negative slacks. As a result, we enable gradients to be computed directly during the AT and slack propagation phases by meticulously managing data dependencies.
\begin{table}[t]
    \centering
    \caption{Statistics of the datasets used in our experiments.}
    \vspace{-1em}
    \label{tab:dataset_stats}
        \resizebox{0.40\textwidth}{!}{
    \begin{tabular}{lrrr}
    \toprule
    Design           & \multicolumn{1}{l}{\#Cells} & \multicolumn{1}{l}{\#Nets} & \multicolumn{1}{l}{\#Pins} \\
    \midrule
    aes\_cipher\_top\cite{ICCADContest2025} & 9,917                       & 10,178                     & 37,357                     \\
    superblue1\cite{2015ICCAD}       & 1,209,716                   & 1,215,710                  & 3,767,494                  \\
    superblue3\cite{2015ICCAD}       & 1,213,252                   & 1,224,979                  & 3,905,321                  \\
    superblue4\cite{2015ICCAD}       & 795,645                     & 802,513                    & 2,497,940                  \\
    superblue5\cite{2015ICCAD}       & 1,086,888                   & 1,100,825                  & 3,246,878                  \\
    superblue7\cite{2015ICCAD}       & 1,931,639                   & 1,933,945                  & 6,372,094                  \\
    superblue10\cite{2015ICCAD}      & 1,876,103                   & 1,898,119                  & 5,560,506                  \\
    superblue16\cite{2015ICCAD}      & 981,559                     & 999,902                    & 3,013,268                  \\
    superblue18\cite{2015ICCAD}      & 768,068                     & 771,542                    & 2,559,143                  \\
    \bottomrule
    \end{tabular}
    }
    \vspace{-1.5em}
\end{table}

We create two separate CUDA streams: one dedicated to STA's core forward (i.e., cell delay and AT propagations) and backward propagation (i.e., slack computations), and another for the differentiable LSE (forward) and its associated gradient computations (backward). This architectural design allows both computational flows to be scheduled concurrently on the GPU.
In such overlapping scheme, the LSE operation for level $i$ must execute only after the completion of the cell delay computation for the same level $i$, as the cell delays derived from LUT interpolations are required for the LSE operation.
Hence, we incorporate CUDA events to manage synchronizations between these streams.
Figure~\ref{fig:gradient_cudastream} depicts the overlapped execution: CUDA events (green blocks) are inserted into the cell delay propagation CUDA stream, and \texttt{cudaStreamWaitEvent} is utilized in the LSE and gradient stream for synchronization. This demonstrates how LSE and gradient computations can be optimized by fusing them with the process of cell AT and slack propagation. Notably, our pin-based parallelization scheme is also applicable to this stage for maximizing parallelism and SIMD lane utilization.

\subsection{Timing-driven Global Placement}
We integrate our GPU STA engine into Xplace3.0~\cite{2025Xplace3}, a timing-driven global placement framework powered by GPU, replacing its original, less efficient GPU-based STA engine~\cite{Wang2014CASTA,Guo2020GPUSTA,Guo2023CPUGPUSTA} as discussed in previous sections. We adopt Xplace3.0's existing pin and path weighting scheme, which optimizes placement based on slacks. In this scheme, each pin is assigned a weight that reflects its criticality. Since slack information is required for all components, the STA engine must be invoked in every placement iteration, leading to significant runtime overhead for large designs.
  \section{Experimental Results and Evaluation}
\label{sec:experiments}

In this section, we evaluate the performance of Warp-STAR against existing CPU and GPU based STA engines: OT~\cite{Huang2021OpenTimerV2} and GPU-Timer~\cite{Wang2014CASTA,Guo2020GPUSTA,Guo2023CPUGPUSTA}, respectively. Next, we analyze the efficiency of gradient computation through \textit{operation fusion}, and compare with INSTA~\cite{lu2025insta}, a GPU-based differentiable STA engine. Finally, we demonstrate the potential of Warp-STAR-based GP framework by comparing our runtime and total negative slacks (TNS) with DREAMPlace 4.0 (DP4.0)~\cite{2023DP4}, Efficient TDP~\cite{2025EfficientTDP}, Xplace3.0~\cite{2025Xplace3} and INSTA-Place~\cite{lu2025insta}.
Our experiments are conducted on a 24-core AMD 7965WX CPU and an NVIDIA RTX 4090 GPU, running with CUDA version 11.8. We use the incremental timing-driven GP contest datasets from the ICCAD 2015~\cite{2015ICCAD} and 2025~\cite{ICCADContest2025}, with statistics in Table~\ref{tab:dataset_stats}.

\begin{table}[t]
    \caption{Runtime comparison (ms.) of Warp-STAR (Ours) with OpenTimer and GPU-Timer.}
    \vspace{-1em}

    \label{tab:timer}
    \centering
    \resizebox{0.45\textwidth}{!}{
    \begin{tabular}{ccccc}
    \toprule
    Design           & \begin{tabular}[c]{@{}c@{}}OpenTimer\\ (OT) \cite{Huang2021OpenTimerV2}\end{tabular} & \begin{tabular}[c]{@{}c@{}}GPU-Timer\\ \cite{Guo2020GPUSTA,Guo2023CPUGPUSTA,Wang2014CASTA}\end{tabular} &  \begin{tabular}[c]{@{}c@{}}Warp-STAR\\(Pin-based)\end{tabular} & \begin{tabular}[c]{@{}c@{}}Warp-STAR\\  (CTE)\end{tabular} \\
    \midrule
    aes\_cipher\_top & 89.1      & 4.03          & 1.49      & 3.96                                                       \\
    superblue1       & 5095      & 57            & 25        & 48                                                         \\
    superblue3       & 5165      & 61            & 26        & 51                                                         \\
    superblue4       & 3149      & 43            & 17        & 31                                                         \\
    superblue5       & 4274      & 48            & 21        & 41                                                         \\
    superblue7       & 8811      & 71            & 35        & 50                                                         \\
    superblue10      & 7551      & 81            & 34        & 71                                                         \\
    superblue16      & 3894      & 40            & 18        & 29                                                         \\
    superblue18      & 3407      & 38            & 16        & 30                                                         \\
    \midrule
    Avg. Speedup     & 1.48\%    & 100.00\%      & 235.97\%  & 124.18\%                                                  \\
    \bottomrule
    \end{tabular}
    }
    \vspace{-1em}
\end{table}
\begin{figure}[t]
    \centering
    \includegraphics[width=0.45\textwidth]{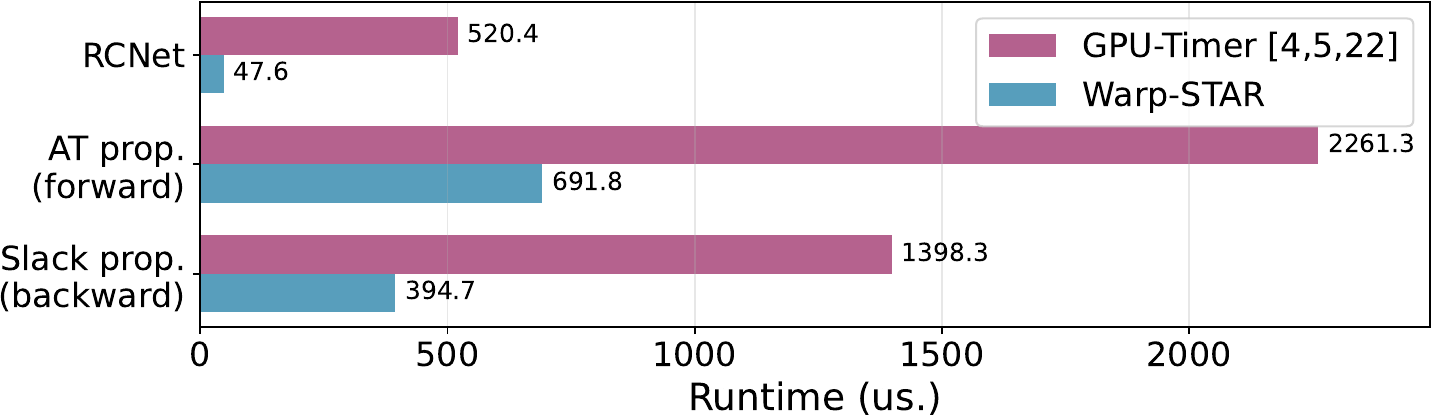}
    \vspace{-1em}
    \caption{Runtime breakdown of the \textit{aes\_cipher\_top} test case.}
    \label{fig:aes_runtime_comparison}
    \vspace{-1.5em}
\end{figure}
\begin{table*}[t]
    \centering
    \caption{Runtime (in seconds) and TNS (in $10^5$ns) of Warp-STAR against existing global placement tools$^1$.}
    \label{tab:gp_comp}
    \vspace{-1em}
    \vbox to 0pt{
        \vss 
        \rlap{
            \fbox{\begin{tabular}{@{}l@{}} 
                \scriptsize $^1$ Both lower runtime \\[-1.0ex] \scriptsize \,\,\, and TNS are better. \\
                \scriptsize $^2$ INSTA-Place uses both \\[-1.0ex] \scriptsize \,\,\, CPU (OT) \& GPU (INSTA).
            \end{tabular}}
        }
        \vspace{-4.8em} 
    }
    \hspace{5em}
    \resizebox{0.75\textwidth}{!}{
    \begin{tabular}{l|rr|rr|rr|rr|rrc}
        \toprule
        \multicolumn{1}{c|}{\multirow{2}{*}{Design}} &
         \multicolumn{2}{|c}{DP4.0~\cite{2023DP4}} &
         \multicolumn{2}{|c}{INSTA-Place~\cite{lu2025insta}} &
         \multicolumn{2}{|c}{Efficient TDP~\cite{2025EfficientTDP}} &
         \multicolumn{2}{|c}{Xplace 3.0~\cite{2025Xplace3}} &
         \multicolumn{3}{|c}{Warp-STAR (\textbf{Ours})} \\
         \multicolumn{1}{c|}{}
                        & TNS    & Time  & TNS    & Time                & TNS    & Time  & TNS    & Time  & TNS    & Time & Time (CTE) \\
         \midrule
         GPU-based STA   & \multicolumn{2}{c|}{\ding{55}} & \multicolumn{2}{c|}{\ding{51}$^2$} & \multicolumn{2}{c|}{\ding{55}} & \multicolumn{2}{c|}{\ding{51}} & \multicolumn{3}{c}{\ding{51}} \\
         \midrule
       aes\_cipher\_top & -1.9   & 6.3   & -      & -                   & -1.9   & 6.5   & \textbf{-1.0}   & 2.8   & \textbf{-1.0}   & \textbf{1.2}  & { }1.6        \\
       superblue1       & -92.3  & 277.0 & -34.6  & \textgreater{}277.0 & \textbf{-15.9}  & 263.8 & -21.8  & 60.9  & -23.7  & \textbf{35.4} & 51.1       \\
       superblue3       & -52.9  & 346.2 & -40.4  & \textgreater{}346.0 & -20.0  & 367.4 & -16.9  & 65.2  & \textbf{-16.8}  & \textbf{40.8} & 57.7       \\
       superblue4       & -151.5 & 143.1 & -114.5 & \textgreater{}143.0 & -89.8  & 540.7 & \textbf{-67.6}  & 42.3  & -78.0  & \textbf{23.5} & 32.2       \\
       superblue5       & -98.5  & 287.0 & -93.7  & \textgreater{}286.0 & -64.4  & 342.0 & -61.7  & 59.0  & \textbf{-61.6}  & \textbf{34.3} & 51.4       \\
       superblue7       & -62.1  & 367.0 & -57.6  & \textgreater{}367.0 & -38.3  & 387.5 & -33.2  & 88.5  & \textbf{-28.9}  & \textbf{58.9} & 70.4       \\
       superblue10      & -674.8 & 472.1 & -628.2 & \textgreater{}472.0 & -562.3 & 783.3 & \textbf{-522.8} & 104.9 & -544.3 & \textbf{59.2} & 91.5       \\
       superblue16      & -67.3  & 189.2 & -37.6  & \textgreater{}189.0 & -24.7  & 184.1 & \textbf{-20.5}  & 42.6  & -23.5  & \textbf{26.3} & 33.0       \\
       superblue18      & -47.3  & 153.6 & -35.8  & \textgreater{}153.0 & -16.0  & 146.4 & -15.2  & 41.8  & \textbf{-15.1}  & \textbf{24.3} & 33.4       \\
       \midrule
       Ratio            & 2.43   & 7.07  & 1.30   & \textgreater{}7.00  & 1.16   & 9.73  & \textbf{0.98}   & 1.74  & 1.0    & \textbf{1.0}  & 1.39  \\
    \bottomrule
    \end{tabular}
    }
    
    \vspace{-1em}
\end{table*}

\subsection{STA Runtime Performance}
Table~\ref{tab:timer} presents the runtime performance of Warp-STAR compared to OT (CPU-based) and GPU-Timer. OT is executed using all 24 CPU cores, while both Warp-STAR and GPU-Timer are executed on a single GPU. Our work presents two implementations: \textit{Warp-STAR (pin-based)} and \textit{Warp-STAR (CTE)}, corresponding to the two methods described in Section~\ref{sec:methodology}. The results show that Warp-STAR achieves an average speedup of 2.4X over GPU-Timer and 162X over OT. The performance improvement is attributed to the elimination of intra-warp load imbalance, which allows for higher SIMD lane utilization on the GPU. Although \textit{Warp-STAR (CTE)} is theoretically more capable of reducing intra-warp load imbalance, it does not show significant improvement over \textit{Warp-STAR (pin-based)}. The CTE method introduces overhead from workload rescheduling and indexing—a limitation also documented by~\cite{2024WER}. Consequently, we adopt the pin-based scheme as our primary implementation.

We further evaluate the runtime performance breakdown of Warp-STAR against GPU-Timer in Figure~\ref{fig:aes_runtime_comparison} for the \textit{aes\_cipher\_top} test case.
Warp-STAR provides a performance boost across all stages due to its increased parallelism and higher SIMD lane utilization.

\begin{table}[t]
    \centering
    \caption{Runtime performance (in ms.) of gradient computation through operation fusion. ``Diff'' refers to sequential gradient computation, ``Diff+Fusion'' refers to the op. fusion.}
    \vspace{-1em}
    \label{tab:diff}
    \resizebox{0.37\textwidth}{!}{
    \begin{tabular}{lccc}
    \toprule
    Design           & Ours  & Ours (Diff) & Ours (Diff+Fusion) \\
    \midrule
    aes\_cipher\_top & 1.49  & 1.72        & 1.61               \\
    superblue1       & 25    & 34          & 29                 \\
    superblue10      & 34    & 47          & 41                 \\
    superblue16      & 18    & 23          & 21                 \\
    superblue18      & 16    & 22          & 19                 \\
    superblue3       & 26    & 36          & 30                 \\
    superblue4       & 17    & 22          & 19                 \\
    superblue5       & 21    & 28          & 24                 \\
    superblue7       & 35    & 46          & 43                 \\
    \midrule
    Norm. Time       & 100\% & 133\%       & 116\%              \\
    \bottomrule
    \end{tabular}
    }
    \vspace{-1.5em}
\end{table}

\subsection{Differentiable STA Runtime Analysis}

\begin{figure}[t]
    \centering
    \includegraphics[width=0.48\textwidth]{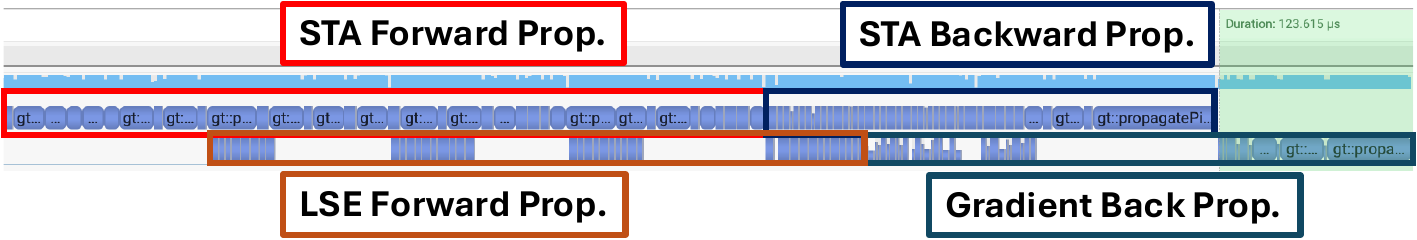}
    \vspace{-2em}
    \caption{Profiling result from NVIDIA Nsight System~\cite{nsys}, illustrating the overlapped execution. The upper part is the CUDA stream for STA engine, while the lower part shows the stream for LSE and gradient computation. We use CUDA events to handle data dependencies every 10 kernel launches.}
    \label{fig:gradient_overlapping}
    \vspace{-1.5em}
\end{figure}

We evaluate the runtime performance of our gradient computation method, which combines our pin-based scheme with the \textit{operation fusion} technique. Table~\ref{tab:diff} presents the runtime performance of gradient computation. The setting ``Diff'' refers to the baseline approach where LSE and gradient computations are performed after STA, while ``Diff+Fusion'' represents our proposed \textit{operation fusion}, which enables concurrent execution with the STA engine. With the employment of optimized warp-level parallelism, the overhead of gradient analysis is only 33\% of the STA engine's runtime. This overhead is further reduced to 16\% with our \textit{operation fusion} technique. Additionally, we profile the actual execution pattern of gradient computation with \textit{operation fusion} in Figure~\ref{fig:gradient_overlapping}. The result depicts that the gradient computation is effectively overlapped with the STA engine, which is the primary reason for the reduced runtime.

We also compare our performance against INSTA. For the \textit{aes-cipher\_top} test case, INSTA takes 108 ms for STA and 61 ms for gradient computation, revealing that gradient computation introduces a 56\% overhead relative to the STA runtime. While INSTA is designed for high accuracy against industrial sign-off tools and is partially closed-sourced, making a direct runtime comparison challenging, our results serve as reference for the performance improvement achievable through our optimization techniques.

\subsection{Timing-driven Global Placement}
To demonstrate the potential of our Warp-STAR-based STA engine, we integrate it into the timing-driven GP framework Xplace3.0~\cite{2025Xplace3} which implements the standard GPU-timer~\cite{Wang2014CASTA,Guo2020GPUSTA,Guo2023CPUGPUSTA}. Our experimental results on the ICCAD 2015 and 2025 contest benchmarks are presented in Table~\ref{tab:gp_comp}.
In terms of execution time, the Warp-STAR-enabled GP framework achieves an average speedup of 7.1X over DP4.0~\cite{2023DP4} and INSTA-Place~\cite{lu2025insta}, 9.7X over Efficient TDP~\cite{2025EfficientTDP}, and 174\% over Xplace3.0~\cite{2025Xplace3}. Due to the substantial overhead of CPU-based STA, DP4.0, INSTA-Place, and Efficient TDP are limited to performing timing analysis every 15 iterations. In contrast, our optimized GPU-based STA is performed in every iteration of our GP flow. Note that since INSTA-Place is not open-sourced, we cite its TNS results directly from its corresponding paper~\cite{lu2025insta}, and its runtime is higher than DP4.0 due to its reliance on the DREAMPlace framework. The experimental results, encompassing both runtime performance and TNS\footnote[3]{Variations in TNS metrics compared to the original Xplace3.0 are attributed to the adoption of a race-free parallel reduction scheme in Warp-STAR, resolving synchronization errors in the baseline GPU-timer and acknowledging the numerical divergence caused by floating-point non-associativity in non-convex optimization.}, demonstrate that Warp-STAR's benefits extend beyond standalone STA, proving its effective applicability to other performance-critical applications like global placement.

  \section{Conclusion}
\label{sec:conc}

In this paper, we present Warp-STAR, a GPU-based STA engine designed to effectively address the overlooked intra-warp load imbalance issue. Incorporating the pin-based scheme, Warp-STAR achieves remarkable performance, demonstrating an average  of 162X speedup over CPU-based approach and 2.4X over GPU-based engines. Furthermore, our \textit{operation fusion} reduces gradient computation overhead to just 16\% of the STA runtime. Finally, integrated into a timing-driven GP framework, Warp-STAR achieves the best runtime costs over SoTA placers, highlighting its broad applicability and performance benefits across timing-driven design flows.
  \begin{acks}
We acknowledge the financial support from Academia Sinica's SiliconMind Project (AS-IAIA-114-M11). This work was also supported in part by the National Science and Technology Council, Taiwan (112-2221-E-002-159-MY3), as well as the National Center for High-performance Computing and Taipei-1 for providing computational resources. We also thank Pei-Che Hsu, Kun-Cheng Wang, and Prof. Hung-Ming Chen from the VDALab at National Yang Ming Chiao Tung University for their assistance during the ICCAD contest.
\end{acks}

  \newpage
  \bibliographystyle{ACM-Reference-Format}
  \bibliography{refs.bib}

  \end{document}